# Scattering correlations of time-gated light

Mykola Kadobianskyi[†], Ioannis N. Papadopoulos[†], Thomas Chaigne[†], Roarke Horstmeyer[†, a], and Benjamin Judkewitz[†, *]

Manipulating the propagation of light through scattering media remains a major challenge for many applications, including astronomy, biomedical imaging and colloidal optics. Wavefront shaping is one of the most promising ways to mitigate scattering and focus through inhomogeneous samples. However, wavefront correction remains accurate over only a limited spatial extent within the scattering medium – a correlation range referred to as the optical memory effect. Here, by selecting only the weakly scattered light for wavefront shaping, we show that the addition of temporal degrees of freedom enhances this correlation range. We investigate spatial scattering correlations by digitally time-gating the early arriving light in the spectral domain. We demonstrate that the range of the translational memory effect for the early arriving light is increased almost fourfold, paving the way for a range of scattering media imaging applications.

[†]Bioimaging and Neurophotonics lab, NeuroCure Cluster of Excellence, Charité – Universitätsmedizin Berlin, Charitéplatz 1, 10117 Berlin, Germany

[a]Future address: Biomedical Engineering Department, Duke University, Durham NC 27708 USA

*corresponding author: benjamin.judkewitz@charite.de

## INTRODUCTION

Due to scattering, high resolution optical imaging of biological tissues is mostly limited to superficial layers. The resolution of fluorescence imaging depends on the ability to form a sharp focus at a plane of interest, which becomes increasingly challenging for tissue layers thicker than a few scattering mean free paths (MFP)[1]. With the advent of wavefront shaping, it became possible to control light propagation and focus both through[2,3] and inside[4-6] turbid media to a diffraction-limited spot, recently demonstrated through highly scattering samples that are almost 10 cm thick[7]. However, this focus spot must be scanned around within the scatterer to form an image. Assuming the plane of interest is at a depth where excitation light arrives after multiple scattering events, different incident wavefronts are needed to focus the incident light into neighboring points. This would require a new round of wavefront correction for every scanned point, making imaging practically impossible for most applications.

Fortunately, first order spatial correlations between the incident and scattered wavefronts can significantly increase the spatial range of accurate single-wavefront corrections. This range is referred to as the memory effect[8], with several related varieties[9]. The angular or "tilt/tilt" memory effect describes the following phenomenon: small tilts of the incident wavefront will result in corresponding tilts of the scattered wavefront at the output surface of a scattering medium. This tilt will then manifest itself as a shift of the resulting speckle pattern at a distant imaging plane. The requirement that the imaging plane is at a distance away from the scatterer renders the tilt/tilt effect of less interest for biological imaging, where scanning must be achieved within tissue at the desired imaging plane. Recently, the translational memory effect – the "shift/shift" memory effect - was discovered and applied for focusing through forward-scattering media[10]. In this case, the shift of the incident wavefront in the transverse plane orthogonal to the propagation direction directly results in a shift of the scattered light speckle pattern at the output plane. The range of this "shift/shift" memory effect was shown to be equal to the average speckle grain size when the incident wavefront is a plane wave, which provides a convenient way to measure the correction range, or field of view (FOV). An optimized incident wavefront can be shifted across the correction FOV along the input plane to scan the optimized focus along the output plane by the same amount.

In biological tissues light scattering is highly anisotropic in the forward direction at visible and near-infrared wavelengths[11]. This means that most of photons traveling through tissue will scatter only in the narrow forward cone (so called "snake" photons[12]). We hypothesize that these snake photons will exhibit a higher level of "shift/shift" correlations. Since the time that the forward-scattered photons spend inside the tissue is shorter due to a shorter optical path, they can be gated by the means of time or coherence gating. This gating can be done either in the time or the frequency domain[12], with the frequency domain being preferable since the spectrum of the time gate can be shaped digitally after the data have been recorded. In this work we study the temporal dependence of speckle correlations for forward-scattering samples and show that time-gating of early arriving photons increases the range of the translational memory effect almost fourfold.

**MATERIALS AND METHODS**

**Time gating of scattered light**

When a short pulse of light propagates through a scattering medium, its temporal envelope elongates and the emergent output pulse can be orders of magnitude longer than the incident one[13]. To better understand the pulse evolution, and to what extent memory effect correlations are maintained for different travel times, we aim to record the time-dependent electric field $E_{out}(r,t)$ at the output surface of the medium. This will allow us to study how different time delays affect the translational memory effect range and by which extent it can be increased for anisotropically scattering media.

One way of implementing coherence gating in the time domain is to employ an interferometric setup (Fig. 1). Using a light source with short temporal coherence, we can reject light outside the coherence window by adjusting the delay between the two interferometer arms[12]. For a pulsed laser under the scalar wave approximation, the field incident upon the scatterer can be expressed as a plane wave

$$E_{inc}(r,t) = u_{inc}(t)e^{ik_0 r}, \qquad (1)$$

where $u_{inc}(t)$ is the temporal envelope of the laser pulse, which can be linked to the envelope of the elongated scattered output field $u_{out}(r,t)$ by introducing a the temporal transfer function of the medium

$$u_{out}(r,t) = \int_0^{+\infty} d\tau\, T_{k_0}(r,t)\, u_{inc}(t-\tau). \qquad (2)$$

Although the transfer function $T_{k_0}(r,\tau)$ depends upon the incident plane wave direction $k_0$, we are primarily interested in its temporal effects for on-axis illumination.

The time gate is implemented by interfering the scattered output field $u_{out}(r,t)$ and a reference pulse. If we introduce an optical path delay $c\Delta t$ between the reference pulse and scattered field and assume on-axis illumination ($k_{0\perp} = 0$), the resulting interferogram recorded by the detector will be

$$I(r, \Delta t) = I_0(r) + \int_0^T dt\ u_{inc}(t - \Delta t)\ u_{out}(t, r) + c.c., \quad (3)$$

where $I_0(r)$ is the non-interfering background, c.c. stands for a similar complex conjugate term and we assume the detector integrates for a time T that is significantly longer than one optical cycle.

We may express the interference term as a convolution $T \star A_{u_{inc}} = \int_0^T dt\ u_{inc}(t - \Delta t)\ u_{out}(r, t)$, where $A_{u_{inc}}(t) = u_{inc} \star u_{inc}$ is the pulse autocorrelation function (or, equivalently, the Fourier transform of its power spectrum). It defines the resolution with which the scattered light can be temporally resolved[14].

An important property can be derived for the scattering medium transfer function T(t) for media with negligible absorption:

$$T(t) \propto \mathcal{F}_{\nu \to t}\{e^{i\Phi(\cdot)}\}(t), \quad (4)$$

where $\Phi$ is a real-valued function. This means that light scattering changes only the spectral phase of the incident pulse's frequency components and not its power spectrum. Thus, the highest temporal resolution for a given power spectrum will be achieved for a bandwidth-limited incident pulse, where all frequency components are in phase. Such a condition is difficult to achieve in the time domain, since coherent broadband sources are hard to implement and dispersion cannot easily be compensated across a wide range of frequencies. This leads us to the necessity of implementing time gating of weakly scattered light in the spectral domain, where we are free to shape the spectrum of the incident pulse as desired.

**Translational memory effect of the scattered light gated in the frequency domain**

A spectral domain approach to time gating was first proposed in[15,16] and recently used to temporally refocus femtosecond pulses through a scattering medium using the spatio-spectral transmission matrix[17]. A laser pulse can be represented as the sum of its monochromatic components through the Fourier transform. Then, if we record a discrete set of speckle patterns at N frequencies $\nu_1..\nu_N$ with an equal step $\Delta\nu$, we can approximate the output wavefront as

$$u_{out}(x, y, t) \approx \Delta\nu \sum_{n=1}^{N} u_{out}(x, y, \nu_n) e^{i2\pi\nu_n t}. \quad (5)$$

The set of scattered monochromatic wavefronts at different frequencies $\{u_{out}(x, y, \nu_n)\}_{n=1..N}$ can be used to reconstruct the scattered wavefronts at different time delays *post hoc* by performing a Fourier transform along the frequency axis of the dataset. Because of discretization, the frequency step $\Delta\nu$ together with the spanned frequency bandwidth will define the time periodicity and the temporal resolution of the synthesized pulse[16].

Due to factors such as dispersion, chromatic focus shift and reference beam phase drift, the recorded monochromatic electric fields are shifted in phase by different unknown global offsets $\{u_{out}(x, y, \nu_n)e^{i\phi(\nu_n)}\}_{n=1...N}$, where $\{\phi(\nu_n)\}_{n=1...N}$ are the unknown spectral phases[18]. When unaccounted for, these global phase shifts essentially stretch the synthesized pulse in time and degrade the temporal resolution. To try to determine them, we start with the assumption that for a shift in frequency much smaller than the spectral bandwidth of the medium[2], change in the speckle pattern can be considered negligible. We can define two discretized reconstructed output fields at neighboring frequencies $u_{kl}(\nu_1) = |u_{kl}(\nu_1)|e^{i\phi_{kl}(\nu_1)}$ and $u_{kl}(\nu_2) = |u_{kl}(\nu_2)|e^{i\phi_{kl}(\nu_2)}$, each M×M pixels in size and with the total intensity normalized to one. If they differ only in global phase shift $\Delta\phi$, we can estimate this phase shift by taking an inner product across all pixels:

$$\sum_{kl} u_{kl}^*(\nu_2)u_{kl}(\nu_1) = \sum_{kl} |u|_{kl}^2 \, e^{-i\Delta\phi} = e^{-i\Delta\phi}. \tag{6}$$

The argument of the inner product on the left side of Eq. 6 gives us the global phase shift between the two reconstructed electric fields. If we compute and then subtract this phase from the dataset in a chain-like manner (between 1st and 2nd, 2nd and 3rd frequency measurement etc.), we can align all of the measured monochromatic electric fields at different frequencies to the same global phase.

We can now summarize our method to measure time-gated scattered fields in the spectral domain (Fig. 2a): we first recorded speckle patterns at a set of equally spaced frequencies, then reconstructed a 3D dataset of scattered electric fields at different frequencies $u_{out}(x, y, \nu)$, then aligned all of the fields to the same global phase via the procedure outlined above and computed the 1D FFT along the frequency dimension of this 3D dataset. The resulting dataset contains the scattered electric fields at the output plane at different time delays, $u_{out}(x, y, t)$.

We note that imprecise alignment and accumulation of spectral phase compensation errors led to a remaining linear phase ramp in the dataset, which caused our reconstructed dataset in the time domain to include a global temporal offset. We manually compensated this offset for each scattering sample in order to realign all output pulses to the same initial time delay, which we define as the arrival of ballistic photons at the output surface.

**Sample preparation**

Most biological tissues exhibit highly anisotropic forward-scattering with a scattering anisotropy factor g > 0.9 (Ref. 11). To mimic the anisotropic behavior of biological tissue we prepared samples of varying thickness consisting of 5 μm silica microspheres (Sigma-Aldrich) in 1% agarose. Using a Mie calculator[19], the scattering properties of the sample were estimated to be g = 0.976 for the anisotropy factor and 90 μm for the scattering MFP.

**Experimental setup**

We used a Mach-Zehnder interferometer to record and extract the scattered complex electric fields at the output surface of the inhomogeneous medium (Fig. 1). The beam from a Ti:Sapphire laser (MaiTai Spectra-Physics) was expanded and attenuated before the interferometric setup (not shown in Figure). The first beamsplitter (BS1) separates the light into two arms: the reference arm and the object arm that illuminates the sample. The combination of the microscope objective OBJ2 (Nikon 40x, NA 0.75) and the tube lens L2 (f = 200 mm) images the scattered light onto an imaging sensor (Basler CMOS2), where it interferes with the reference beam in an off-axis configuration. The optical paths of the reference and object arms were matched before every experiment to ensure the highest fringe contrast by tuning the reference arm length and observing the interference intensity with the laser in pulsed mode. After switching the laser source to continuous wave (CW) mode, we varied the wavelength output from 690 to 940 nm with 801 equally spaced frequency steps and recorded a set of interferograms.

Scattered electric fields were reconstructed for each frequency and normalized to unitary power. Weighting different frequencies allows us to shape the spectrum of the illumination source together with its temporal profile. We chose a uniform spectrum shape resulting in a sinc temporal profile with a time step of ~ 9 fs and a time period of ~ 7 ps. According to Eq. 5, a Fourier transform along the frequency axis produced an electric field at different time delays.

Using the findings in Ref. 10, we measured the extent of the translational memory effect for different output wavefront delay times in two independent ways (see inset on Fig. 1). First, we illuminated the sample with a plane wave projected onto the input surface of the sample by the combination of tube lens L1 and microscope objective OBJ1 (Nikon 20x, NA 0.75) and reconstructed the electric field at different time delays as described above. The range of the translational memory effect was then calculated from the mean speckle grain size at the output surface of the scattering medium, defined as the full width at half maximum (FWHM) of the speckle autocorrelation. In a second experiment, we focused light onto the input surface of the sample and then manually translated the sample while measuring the decorrelation of the output field.

**RESULTS AND DISCUSSION**

Starting with a 360 μm thick sample (equal to 4 scattering MFP), we reconstructed the temporal evolution of the speckle patterns at the output surface of the scattering medium in response to plane wave illumination, as shown in Fig. 2b. The speckle pattern changed as a function of time, with the size of the speckle grains decreasing with increasing time delay. The average speckle size across different time positions was approximately equal the average speckle grain size produced by CW illumination. After performing a 2D spatial Fourier transform of the resulting speckle patterns at different time delays, we observed a ring-like spread of k-vectors that grows as a function of time (Fig. 2c). This indicates that later-arriving scattered photons emerged from the medium at increasingly larger angles. The experiment was repeated for two thicker samples of 720 and 1080 μm (8 MFP and 12 MFP, respectively), and the results of the speckle autocorrelation at different time delays are summarized in Fig. 3a. The mean speckle size was the largest at the earliest time delay and quickly decreased at later time delays (Fig. 3a). For the three different samples we compared the speckle autocorrelation of the early-arriving scattered light with that of CW illumination at a central wavelength of 815 nm (Fig. 3b). In each sample we observed that the earlier-arriving photons exhibit a wider range of translational memory effect compared to the CW case, as manifested by a larger mean speckle size for plane wave illumination. For the thinnest sample, the earliest time delay gave almost a fourfold increase in the memory effect range, while for thicker samples, where ballistic light is virtually absent, it was still more than twofold. Since the synthesis of the pulse was done digitally in the frequency domain, we could select a different bandwidth of the excitation source and therefore create different time gate durations. Comparing the results of the memory effect extent as a function of excitation bandwidth (Fig. 4a), we observed that the mean speckle size (the range of the translational memory effect) increased with larger bandwidths (i.e., with narrower time gate durations).

For the second measurement, we formed a focus spot on the input plane of the scattering sample by removing the tube lens L1 and again recorded a set of electric fields at different frequencies. The focus was positioned at the input plane of the scattering medium with the help of a reflection imaging system (tube lens L3 and Basler CMOS1) placed before the focusing objective OBJ1. We then physically translated the sample in $\Delta x = 0.5$ μm steps, perpendicular to the incident beam direction and relative to the input focus spot, and repeated the measurement for a total translation distance of 4 μm. The shift/shift correlation function $C_{shift/shift}(n\Delta x)$ was obtained as the absolute value of the cross-correlation between the complex scattered electric fields at two positions with a relative distance of $n\Delta x$.

Since the shift/shift correlation function only depends on the relative distance between any two points and not their absolute positions, we first created a m x m matrix of pairwise correlation coefficients $\|C_{ij}\|$, where $C_{ij}$ is the absolute value of the correlation coefficient between the electric fields at shift positions $i\Delta x$ and $j\Delta x$. The final shift/shift correlation function value at a shift $n\Delta x$ can then be calculated as the average along the corresponding side diagonal of the pairwise correlation matrix:

$$C_{shift/shift}(n\Delta x) = \frac{1}{m-n+1} \sum_{i-j=n} C_{ij}, n = 1\ldots m. \tag{7}$$

Calculating the correlation coefficient with Eq. (7) allows for more pairwise correlation coefficients to be averaged together for shorter shifts (position closer to the main diagonal of the matrix), therefore offering a better approximation to the shape of the shift/shift correlation function, especially at its onset.

We compared the two independent measurements of the translational memory effect range for the thinnest sample, both for the earliest arriving time-gated light and the response under CW excitation (center wavelength of 815 nm), in Fig 4b. Our two independent measurements of the memory effect range, via speckle autocorrelation and via sample shifting, both matched and also both confirmed that earlier arriving light exhibits a larger translational FOV.

**CONCLUSIONS**

Controlling light propagation and shaping the incident wavefront to image through thick scattering media is time consuming. In anisotropically scattering media such as biological tissue, the translational memory effect allows to use the same wavefront to focus and then scan within the small area around the original focus simply by shifting the wavefront pattern at the input plane.

Forward-scattered photons that are deflected at small angles from their original direction are expected to exhibit a longer "shift/shift" correlation scan range. However, under CW illumination those photons are mixed with diffuse light and their properties cannot be exploited. In this study we digitally synthesized an ultra-short pulse from a set of monochromatic holograms recorded at different frequencies. Broadband illumination allowed us to reliably gate only the early-arriving photons and demonstrate that the range of the translational memory effect that those photons exhibit is almost fourfold larger than the average effect under CW illumination. The FOV increase was clearly observed for all studied scattering samples, with the FOV decreasing slowly as the sample thickness increased and the proportion of weakly scattered light diminished.

For this study we employed most of the laser bandwidth that could be addressed for uninterrupted wavelength tuning. As demonstrated by the dependency of the translational memory effect FOV on the illumination bandwidth (Fig. 4a), a wide bandwidth was essential for achieving high enough temporal resolution to reliably select only early-arriving light. This can be explained by the fact that highly anisotropic forward-scattering media have short confinement times and the emergent output pulse is relatively weakly elongated compared to the incident pulse.

Time-gating of early-arriving "snake" photons with wide bandwidth sources could be combined with non-linear deep tissue imaging methods that rely on wavefront correction to focus through scattering tissue layers[20-22]. These techniques indirectly select the first-arriving light through iterative wavefront optimization. Adding the temporal degree of freedom in wavefront shaping can enable a more precise control of light propagation through scattering media, therefore paving the way towards more efficient scattering compensation for deep tissue imaging.

**ACKNOWLEDGMENTS**

We would like to thank M. Mounaix for technical advice and M. Hoffmann for helpful discussions. This work was supported by the European Research Council (ERC- 2016-StG- 714560), Einstein Foundation Berlin, DFG (EXC 257 NeuroCure) and the Krupp Foundation.

**REFERENCES**

1    Ntziachristos V. Going deeper than microscopy: the optical imaging frontier in biology. *Nature Methods* 2010; **7**: 603–614.
2    Mosk AP, Lagendijk A, Lerosey G, Fink M. Controlling waves in space and time for imaging and focusing in complex media. *Nature Photonics* 2012; **6**: 283–292.
3    Yaqoob Z, Psaltis D, Feld MS, Yang C. Optical phase conjugation for turbidity suppression in biological samples. *Nature Photonics* 2008; **2**: 110–115.
4    Xu X, Liu H, Wang LV. Time-reversed ultrasonically encoded optical focusing into scattering media. *Nature Photonics* 2011; **5**: 154–157.
5    Wang YM, Judkewitz B, DiMarzio CA, Yang C. Deep-tissue focal fluorescence imaging with digitally time-reversed ultrasound-encoded light. *Nature Communications* 2012; **3**: 928.
6    Horstmeyer R, Ruan H, Yang C. Guidestar-assisted wavefront-shaping methods for focusing light into biological tissue. *Nature Photonics* 2015; **9**: 563–571.
7    Shen Y, Liu Y, Ma C, Wang LV. Focusing light through biological tissue and tissue-mimicking phantoms up to 9.6 cm in thickness with digital optical phase conjugation. *Journal of Biomedical Optics* 2016; **21**: 085001–9.
8    Feng S, Kane C, Lee PA, Stone AD. Correlations and Fluctuations of Coherent Wave Transmission through Disordered Media. *Physical Review Letters* 1988; **61**: 834–837.
9    Osnabrugge G, Horstmeyer R, Papadopoulos IN, Judkewitz B, Vellekoop IM. The generalized optical memory effect. *Optica* 2017, *in press*.
10   Judkewitz B, Horstmeyer R, Vellekoop IM, Papadopoulos IN, Yang C. Translation correlations in anisotropically scattering media. *Nature Physics* 2015; **11**: 684–689.
11   Cheong WF, Prahl SA, Welch AJ. A review of the optical properties of biological tissues. *IEEE Journal of Quantum Electronics* 1990; **26**: 2166–2185.
12   Dunsby C, French PMW. Techniques for depth-resolved imaging through turbid media including coherence-gated imaging. *Journal of Physics D: Applied Physics* 2003; **36**: R207–R227.
13   Hee MR, Izatt JA, Jacobson JM, Fujimoto JG, Swanson EA. Femtosecond transillumination optical coherence tomography. *Optics Letters* 1993; **18**: 950–952.
14   Schmitt JM. Optical coherence tomography (OCT): a review. *IEEE Journal of selected topics in quantum electronics* 1999; **5**: 1205–1215.
15   Arons E, Dilworth D, Shih M, Sun PC. Use of Fourier synthesis holography to image through inhomogeneities. *Optics Letters* 1993; **18**: 1852–1854.
16   Arons E, Dilworth D. Analysis of Fourier synthesis holography for imaging through scattering materials. *Applied Optics* 1995; **34**: 1841–1847.
17   Mounaix M, Defienne H, Gigan S. Deterministic light focusing in space and time through multiple scattering media with a time-resolved transmission matrix approach. *Physical Review A* 2016; **94**: 041802–5.
18 Andreoli D, Volpe G, Popoff S, Katz O, Grésillon S, Gigan S. Deterministic control of broadband light through a multiply scattering medium via the multispectral transmission matrix. *Scientific Reports* 2015; **5**(10347):1–8.
19 Mätzler C. MATLAB functions for Mie scattering and absorption, version 2. *IAP Res Rep* 2002.
20 Tang J, Germain RN, Cui M. Superpenetration optical microscopy by iterative multiphoton adaptive compensation technique. *Proc Natl Acad Sci USA* 2012; **109**: 8434–8439.
21 Wang C, Liu R, Milkie DE, Sun W, Tan Z, Kerlin A *et al.* Multiplexed aberration measurement for deep tissue imaging in vivo. *Nature Methods* 2014; **11**: 1037–1040.
22 Papadopoulos IN, Jouhanneau J-SB, Poulet JFA, Judkewitz B. Scattering compensation by focus scanning holographic aberration probing (F-SHARP). *Nature Photonics* 2016; **11**: 116–123.

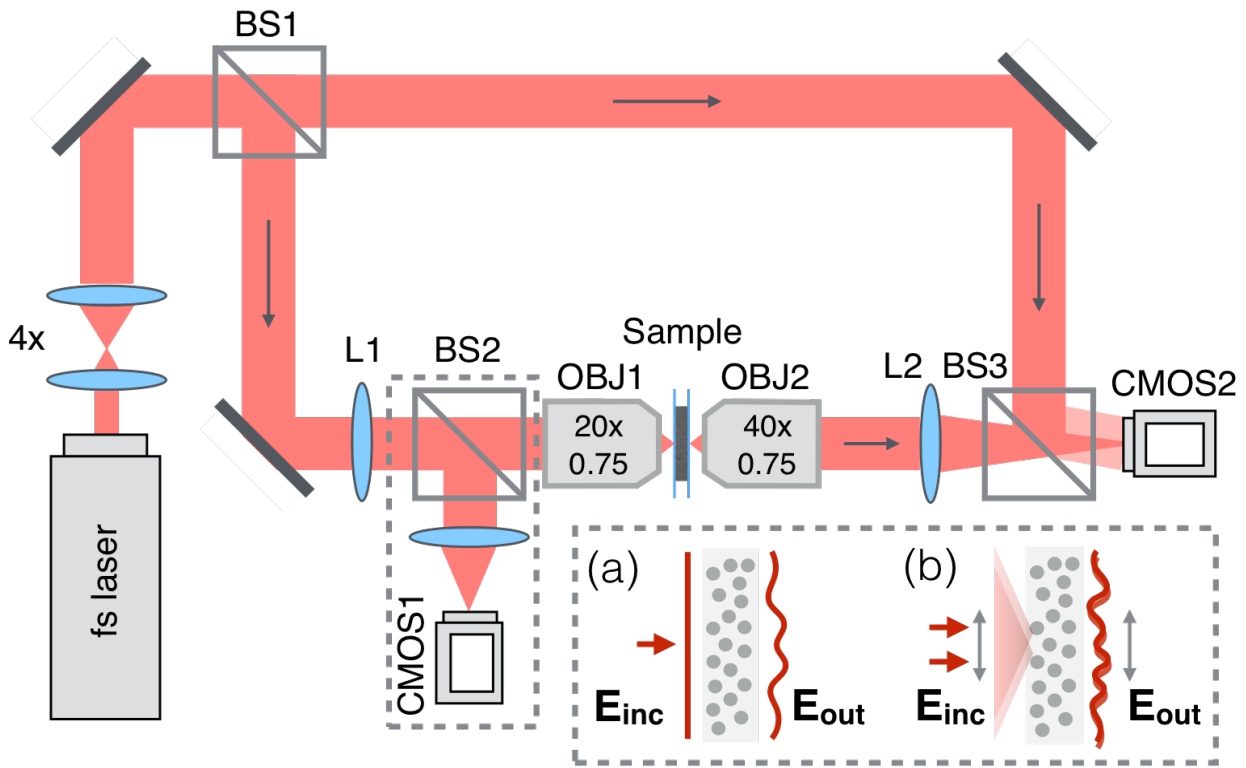

**Figure 1. Principal optical setup.** The output of a tunable laser source is split into two arms of a Mach-Zehnder interferometer by the beamsplitter BS1. The combination of lens L1 and microscope objective OBJ1 projects a plane wave at the input plane of the scattering medium. A focus spot can be formed and precisely positioned at the same plane by removing L1 and focusing using a reflection imaging system (lens L3 and CMOS1). The scattered light at the output plane of the medium is imaged by the transmission imaging system (OBJ2 and L2) onto the CMOS2 where it is combined with the reference beam forming a digital hologram. Inset: The translational memory effect can be measured in two ways: (a) as the width of speckle autocorrelation resulting from plane wave input and (b) as the cross-correlation coefficient between fields produced by physically translating the sample under input focal spot illumination.

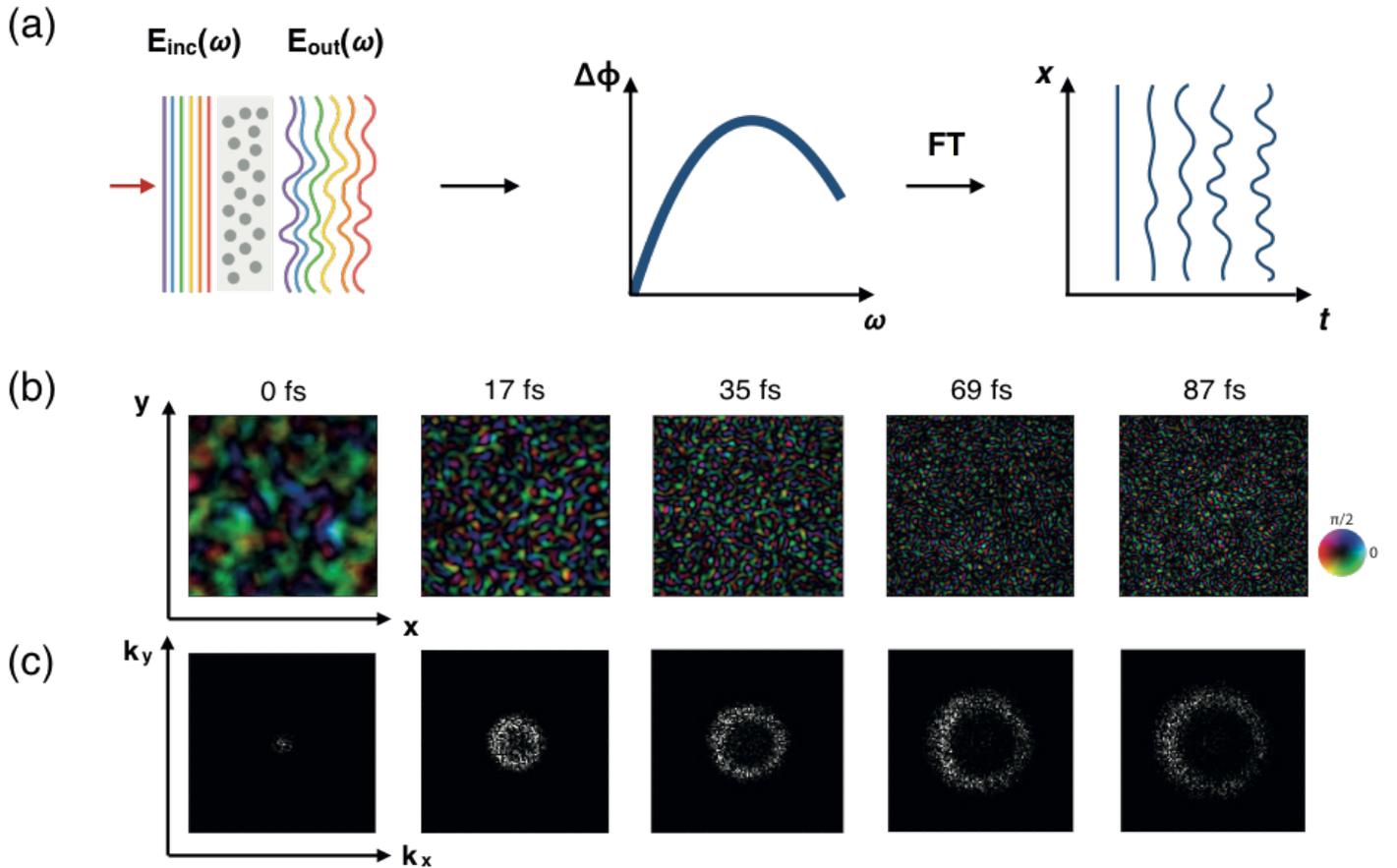

**Figure 2**. **Time gating of the scattered light in the spectral domain. (a)** Sketch of digitally time-gated electric field reconstruction. The amplitudes and phases of the complex speckle pattern at the output of a scattering medium are measured for a set of equally spaced frequencies (left), which are then brought to the same global phase (middle). The scatterer's temporal response is reconstructed by taking a Fourier transform across the frequency dimension (right). **(b)** Amplitude and phase (color) of the complex speckle pattern at different time delays for the thinnest sample (field of view 44 by 44 μm). **(c)** Spatial Fourier transform of each resulting speckle patterns. Photons at larger time delays emerge from the medium at an increasing angle.

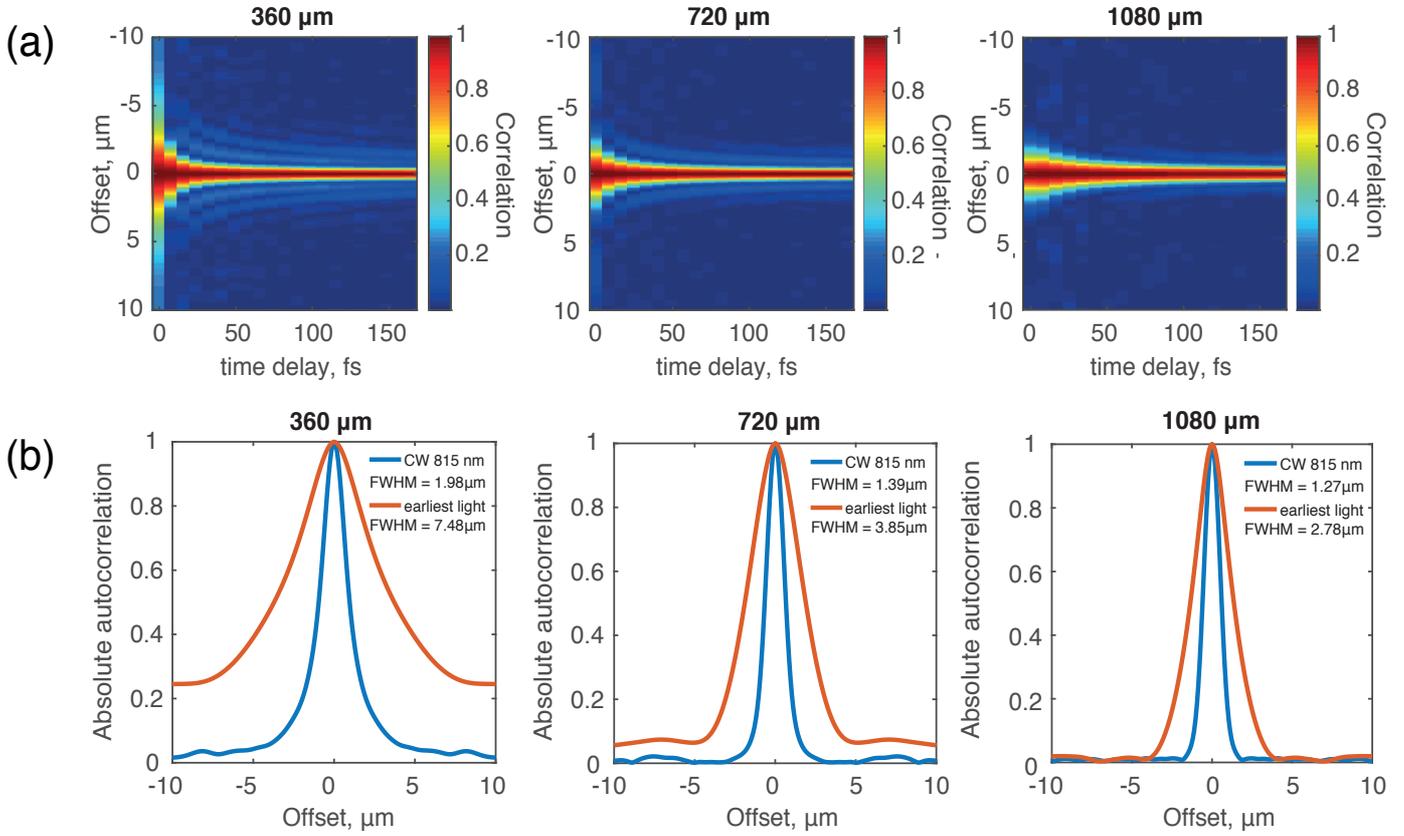

**Figure 3. Speckle autocorrelation of digitally time-gated light. (a)** Speckle autocorrelation as a function of time delay for three samples (under plane wave illumination) with varying thickness, 360 μm (left), 720 μm (center) and 1080 μm (right). **(b)** Speckle autocorrelation for the earliest time delay compared to the non-gated speckle (CW illumination, 815 nm). The translational memory effect range is wider for earlier-arriving light as compared to CW illumination.

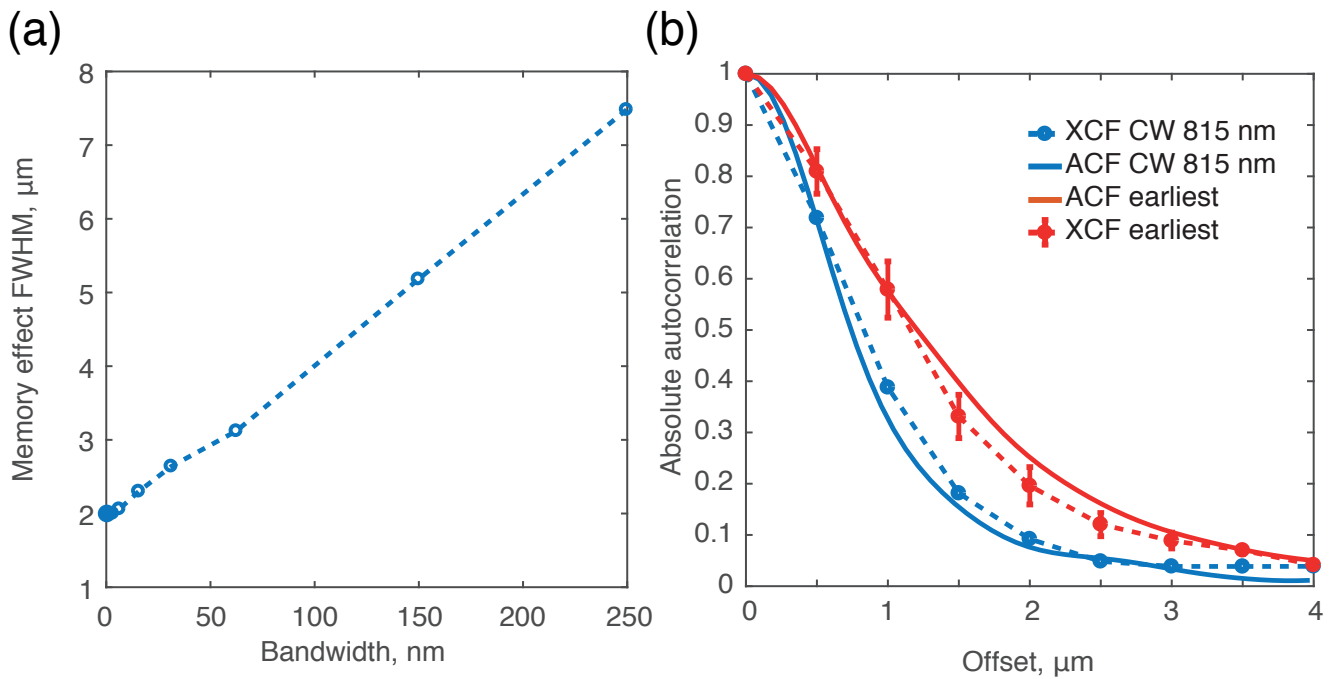

**Figure 4. Influence of bandwidth on the translational memory effect range and comparison of this range for early-arriving "snake" photons vs. CW response based on two independent measurements.** (a) Dependence of the speckle autocorrelation FWHM (mean speckle size) on the bandwidth of the synthesized pulse for the thinnest sample. A short (broadband) pulse is essential for observing an increased memory effect for early-arriving snake photons. **(b)** Speckle autocorrelation value (ACF) measured for an input plane wave illumination (solid lines) and the "shift/shift" memory effect (XCF) correlation value measured for an input focus (dotted lines). Here we show ACF and XCF measurements for both CW illumination (central wavelength 815 nm, blue curves) and for digitally gated early-arriving "snake" photons (red curves). Both measurements indicate an increase in the FOV of the translational memory effect. Data for the early-arriving "snake" photons presented as mean ± standard error of the mean (SEM).